\documentclass[conference]{IEEEtran}
\usepackage[table,xcdraw,usenames,dvipsnames]{xcolor}
\usepackage{array}
\usepackage{float}
\usepackage{graphicx}
\usepackage[english]{babel}
\usepackage{pgfplots}
\usepackage{amsmath}
\usepackage{booktabs}
\usepackage{multirow}
\usepackage{array,booktabs,arydshln,xcolor}

\usepackage{tikz}
\usepackage[table,xcdraw]{}
\usepackage{float}

\begin{document}

\title{Workload Trace Generation for Dynamic Environments in Cloud Computing}
\author{\IEEEauthorblockN{Jammily Ortigoza}
\IEEEauthorblockA{Science and Technology School\\Catholic University of Asunci\'on\\jortigozaf@gmail.com\\Paraguay}
\and
\IEEEauthorblockN{Fabio L\'opez-Pires}
\IEEEauthorblockA{Itaipu Technological Park\\National University of Asunci\'on\\fabio.lopez@pti.org.py\\Paraguay}
\and
\IEEEauthorblockN{Benjam\'in Bar\'an}
\IEEEauthorblockA{National University of Asunci\'on\\Catholic University of Asunci\'on\\bbaran@pol.una.py\\Paraguay}
}

\maketitle

\begin{abstract}
Cloud computing datacenters provide millions of virtual machines in actual cloud markets. In this context, Virtual Machine Placement (VMP) is one of the most challenging problems in cloud infrastructure management, considering the large number of possible optimization criteria and different formulations that could be studied. Considering the on-demand model of cloud computing, the VMP problem should be optimized dynamically to efficiently attend typical workload of modern applications. This work presents possible classification criteria for different formulations of the VMP problem from the Cloud Service Providers' (CSPs) perspective in dynamic environments, based on the most relevant dynamic parameters studied so far in the VMP literature. Several examples for understanding the possible dynamic environments are presented for future implementation of workload trace generation to deeply studies and further advance on this research area. Finally, future directions are also presented.
\end{abstract}
\IEEEpeerreviewmaketitle

\section{Introduction}
{\setlength{\parindent}{12pt}	

The rapid demand growth for computational resources in modern business and scientific applications represents several challenges for design, implementation and management of scalable datacenters to meet the requirements of customers in a competitive and efficient way \cite{soundararajan2010challenges}. 

Considering the evolution of resource provisioning, three main models could be identified: (1) traditional provisioning of resources with independent physical hardware, (2) modern provisioning of shared resources through virtualized hardware and (3) trending dynamic provisioning of resources through a cloud computing model \cite{buyya2008market}. The traditional provisioning environment has mostly evolved to a virtualized provisioning of resources in current datacenters, considering its advantages for management and resource utilization.

Virtualization in modern datacenters introduces complex management decisions related to the placement of virtual machines (VMs) into the available physical machines (PMs). In this context, Virtual Machine Placement (VMP) represents the process of selecting which VMs should be executed in a given set of PMs of a datacenter \cite{lopez2015b}. The VMP problem is mostly formulated as a combinatorial optimization problem, representing one of the most challenging problems in virtualized datacenters infrastructure management, considering the large number of possible optimization criteria and different formulations that could be studied \cite{lopez2015}.

For virtualized datacenters with deployments of VMs that rarely change its configuration over time, a static (offline) formulation of the VMP problem may be appropriate \cite{lopez2013virtual}. Additionally, in virtualized datacenters where a small number of VMs are created and destroyed, a semi-static formulation of the VMP problem could be acceptable (e.g. consolidating VMs every day at midnight). On the other hand, considering the today more realistic on-demand model of cloud computing with dynamic resource provisioning, a static (or semi-static) formulation of the VMP problem can result in under-optimal solutions after a short period of time. Clearly, the VMP problem for cloud computing environments must be formulated as a pure dynamic (online) optimization problem to efficiently attend dynamic workload of modern applications \cite{lopez2015}.

\subsection{Background and Motivation}
\label{motivation}
The VMP problem has been extensively studied and several surveys have already been presented in the VMP literature. Existing surveys focus on specific issues such as: (1) energy-efficient techniques applied to the problem \cite{beloglazov2012energy,salimian2013survey}, (2) particular architectures where the VMP problem is applied, as federated clouds \cite{gahlawat2014survey}, and (3) methods for comparing performance of placement algorithms in large on-demand clouds \cite{mills2011comparing}. None of the mentioned surveys presented a general and extensive study of a large part of the VMP literature. In consequence, L{\'o}pez-Pires and Bar{\'a}n presented in \cite{lopez2015} an extensive up-to-date survey of the most relevant VMP literature and proposed a novel taxonomy in order to identify research opportunities defining a general vision on this problem.

According to \cite{lopez2015}, the VMP problem is mostly formulated as an online optimization problem, where live migration techniques allow VMs to be dynamically consolidated on necessary PMs according to dynamic requirements of resources. The most studied environment for online formulations of the VMP problem considers that VMs are dynamically created and destroyed \cite{lopez2015}. To the best of the authors' knowledge, there is no published work presenting a detailed characterization of possible dynamic environments for the VMP problem.

Clearly, a deeper research of possible dynamic parameters in cloud computing is necessary in order to propose holistic and more realistic environments for the formulation of the VMP problem for cloud computing datacenters. 

Consequently, it is important to extend the taxonomy presented in \cite{lopez2015} focusing on dynamic formulations of the VMP problem from the providers' perspective, in order to understand possible challenges for Cloud Service Providers (CSPs) in dynamic environments to efficiently attend customers' requests for virtual resources, based on the most relevant dynamic parameters studied so far in the VMP literature.

\section{Reviewed Literature}
\label{literature}

\subsection{Keywords Search}
The selection process of relevant articles started with a search for research articles from Google Scholar database [scholar.google.com] with at least one of the following selected keywords in the article title: (1) virtual machine placement, (2) vm placement, (3) virtual machine consolidation, (4) vm consolidation or (5) server consolidation. The search of those keywords resulted in 446 research articles. A detailed list of the 446 resulting articles can be found in \cite{lopez2014survey}.

\subsection{Publisher Filtering}
Considering the large number of results from keywords search step, the literature selection process focused on research articles from the following well-known publishers: (1) ACM, (2) IEEE, (3) Elsevier and (4) Springer. This filtering step resulted in a reduction from 446 to 172 research articles. A detailed list of the 172 resulting articles can be found in \cite{lopez2014survey}.

\subsection{Abstract Reading}
Considering the 172 resulting articles from the publisher filtering step, a reading of the abstracts was performed in order to identify the most relevant articles that specifically study the VMP problem. Additionally, short papers (i.e. research articles with less than 6 pages) were removed from the selected literature, resulting in 84 selected articles of the VMP literature. A detailed list of the 84 resulting articles can be found in \cite{lopez2014survey}.

\subsection{Online Formulations for Provider-oriented VMP Problem}
Based on the  84 studied articles from \cite{lopez2015}, this work focuses on the 64 articles that proposed online formulations for the VMP problem from the providers' perspective, considering the relevance of this type of environments for actual cloud computing providers. An in-depth reading of this universe of 64 articles was performed with the aim of identifying the most relevant dynamic parameters.

\section{Classification Criteria}
\label{criteria}
This work identified the following dynamic parameters: 

\begin{itemize}	
\item resource capacities of VMs (vertical elasticity);
\item number of VMs of a service (horizontal elasticity);
\item utilization of resources of VMs (related to overbooking).
\end{itemize}

Consequently, dynamic environments for online formulations of the provider-oriented VMP problem may be classified by one or more of the following classification criteria: \hspace{1cm} (1) elasticity and (2) overbooking, as presented in the following subsections.

\begin{figure}[b]
\centering
\includegraphics[scale=0.45]{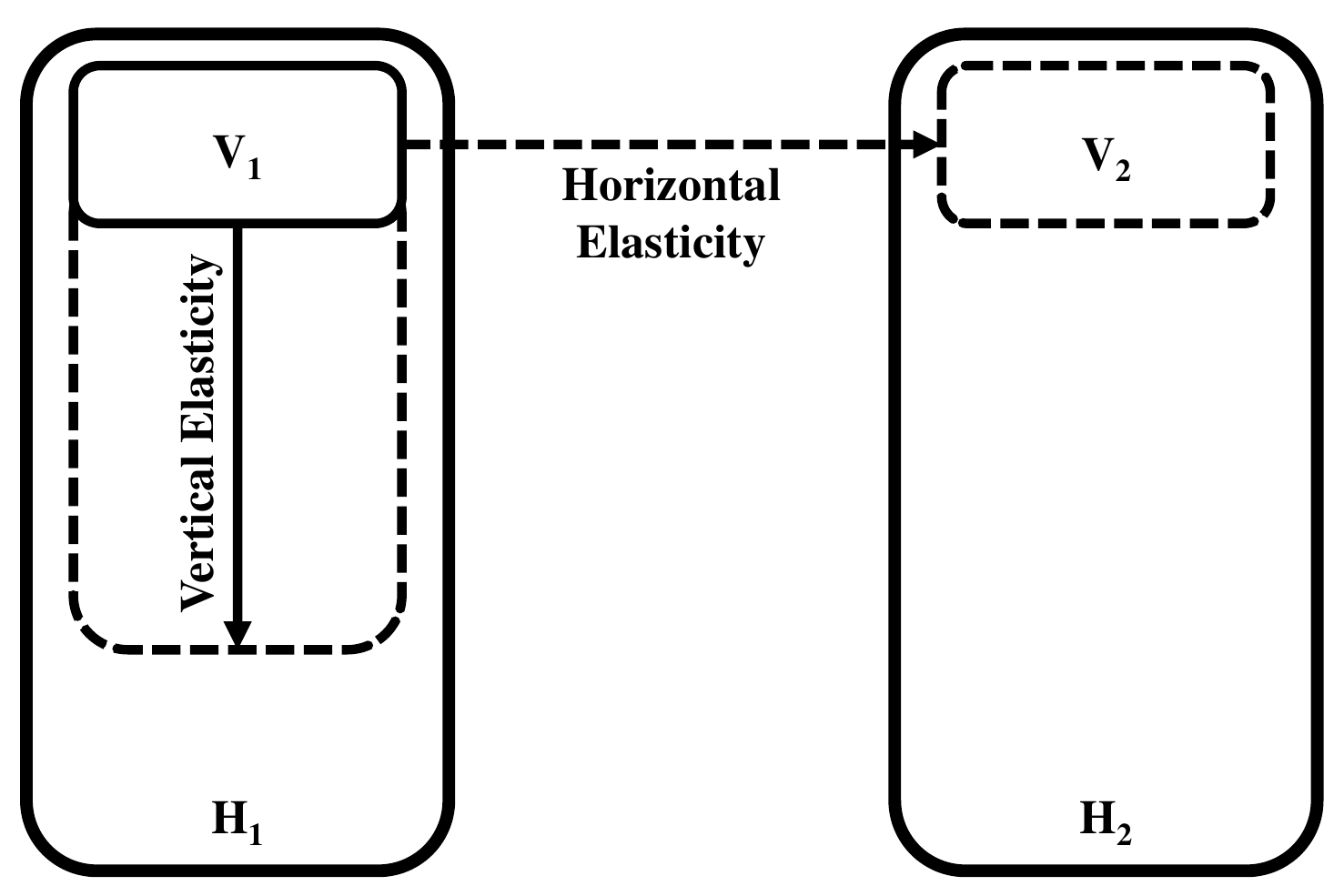}
\caption{Vertical and horizontal elasticity. Vertical elasticity dynamically adjusts the capacities of virtual resources inside a VM while horizontal elasticity dynamically adjusts the number of VMs (e.g. in distributed applications).}
\label{fig:elasticity}
\end{figure}

\subsection{Elasticity}
\label{sec:elasticity}
According to the definition given in \cite{6211900}: ``Cloud elasticity is the ability of a cloud infrastructure to rapidly change the amount of resources allocated to a service in order to meet the actual varying demands on the service while enforcing a Service Level Agreement (SLA)". 

Considering the dynamic workload of modern cloud applications, proactive elasticity is a very important issue to address for CSPs in order to deal with under-provisioning (saturation) and over-provisioning (under-utilization) of cloud resources \cite{armbrust2009m}. Under-provisioning can cause SLA violations, impacting directly on economical revenue while over-provisioning can cause inefficient utilization of resources, directly impacting on resource utilization and energy consumption.

Research articles considering online formulations of the provider-oriented VMP literature have already studied two types of elasticity: vertical and horizontal (see Figure \ref{fig:elasticity}). Vertical elasticity can be defined as the ability of cloud services to dynamically change capacities of virtual resources (e.g. CPU and RAM memory) inside a VM, while horizontal elasticity can be defined as the ability of cloud services to dynamically adjust the number of VMs associated to a cloud service \cite{Wang2012}.

Implementing vertical elasticity requires shorter time of service reconfiguration than horizontal elasticity, but with a higher migration cost. On the other hand, horizontal elasticity enables stronger high availability than vertical elasticity, but a coordination overhead is required and infrastructure complexity increases \cite{Wang2012}.

\subsection{Overbooking}
\label{sec:overbooking}
Resources overbooking can make cloud services more profitable for CSPs, overlaying requested virtual resources onto physical resources at a higher ratio than 1:1 \cite{hoeflin2012quantifying}. Online formulations of the provider-oriented VMP considering overbooking include particular considerations to efficiently attend customers' requirements, enforcing SLAs. 

Considering the dynamic workload of cloud applications and services, virtual resources of VMs are also dynamically used giving space to re-utilization of idle resources that were already reserved. Research articles considering online formulations of the provider-oriented VMP literature have already studied two types of overbooking: server and network resources overbooking. 

In this context, CSPs should measure the utilization of resources of VMs in order to correctly manage the overbooking with the available physical resources, minimizing SLA violations. Monitoring utilization of virtual resources and workload of cloud applications and services also helps CSPs to consider forecasting techniques for approximating in advance the required management actions (e.g. migrations of VMs) for the consolidation process, to reduce resource under-provisioning \cite{fang2013vmplanner,zhang2014dynamic}.

\section{Possible Classifications}
\label{taxonomy}
Based on the universe of 64 studied research articles, dynamic environments for online formulations of the provider-oriented VMP problem may be classified by one or more of the following classification criteria: (1) elasticity and (2) overbooking, as presented in Section \ref{criteria}. 

First, dynamic environments could be formulated considering one of the following elasticity values: 

\begin{itemize}	
\item elasticity=0: no elasticity;
\item elasticity=1: horizontal elasticity;
\item elasticity=2: vertical elasticity;
\item elasticity=3: horizontal and vertical elasticity. 
\end{itemize}

Additionally, identified dynamic environments may also consider one of the following overbooking values: 

\begin{itemize}
\item overbooking=0: no overbooking;
\item overbooking=1: server resources overbooking;
\item overbooking=2: network resources overbooking;
\item overbooking=3: server and network overbooking.
\end{itemize}

Based on the combinations of the possible values of the classification criteria (elasticity,overbooking), it has been identified 16 different possible environments. Considering this representation, each identified dynamic environment is denoted by its elasticity and overbooking coordinates. 

\subsection{Cloud Service and Environment Notation}
\label{cs_notation}
CSPs dynamically receive requests for the placement of cloud services with different characteristics according to the classification criteria presented in Section \ref{criteria}, representing real-world environments and generalizing the deployment of cloud services in several possible cloud architectures (e.g. single-cloud, distributed-cloud or federated-cloud). Cloud services may represent simple services such as Domain Name Service (DNS) or complex multi-tier elastic applications.

A cloud service is composed by a set of VMs, where each VM of a cloud service could be located for its execution in different cloud datacenters according to the customers preferences or requirements (e.g. legal issues or high-availability). 

Configuration of VMs of a cloud service changes dynamically when elasticity is considered. On the other hand, utilization of virtual resources change dynamically according to the demand when overbooking is considered; otherwise, the utilization of each virtual resource is considered at 100\%.

Formally, a cloud service $S_{b}$ can be distributed across different possible cloud datacenters. Each cloud datacenter $DC_c$ hosts VMs $V'_{cj}$ associated to different cloud services. A VM $V'_{cj}$ associated to a service $S_{b}$ is denoted as $V''_{bcj}$.
\\\\
\noindent where:

\begin{tabbing}
\hspace*{1.65cm}\=\kill
	$S_{b}$:	\>Cloud service $b$;\\
	$DC_{c}$: \>Cloud datacenter $c$;\\
	$mDC_{c}$: \>Number of VMs $V_{j}$ in $DC_{c}$;\\	
	$mS_{b}$: \>Number of VMs $V_{j}$ in $S_{b}$;\\	
	$V'_{cj}$:	\>$V_j$ in $DC_{c}$;\\
	$V''_{bcj}$:\>$V_j$ in $DC_{c}$ from service $S_{b}$.\\
\end{tabbing}

Figure \ref{fig:service} presents a basic example of a cloud service $S_1$, distributed across 2 cloud datacenters $DC_1$ and $DC_2$ and using 4 VMs: $V''_{111}$, $V''_{112}$, $V''_{121}$, $V''_{122}$. These cloud datacenters could represent geo-distributed datacenters owned by one CSP or a federated-cloud with two different CSPs. Each cloud datacenter hosts 2 VMs of $S_1$: $V'_{11}$ and $V'_{12}$ represent $V_{1}$ and $V_{2}$ in $DC_1$ respectively (analogously $DC_2$ hosts 2 VMs).

\begin{figure}[b]
\centering
\includegraphics[scale=0.5]{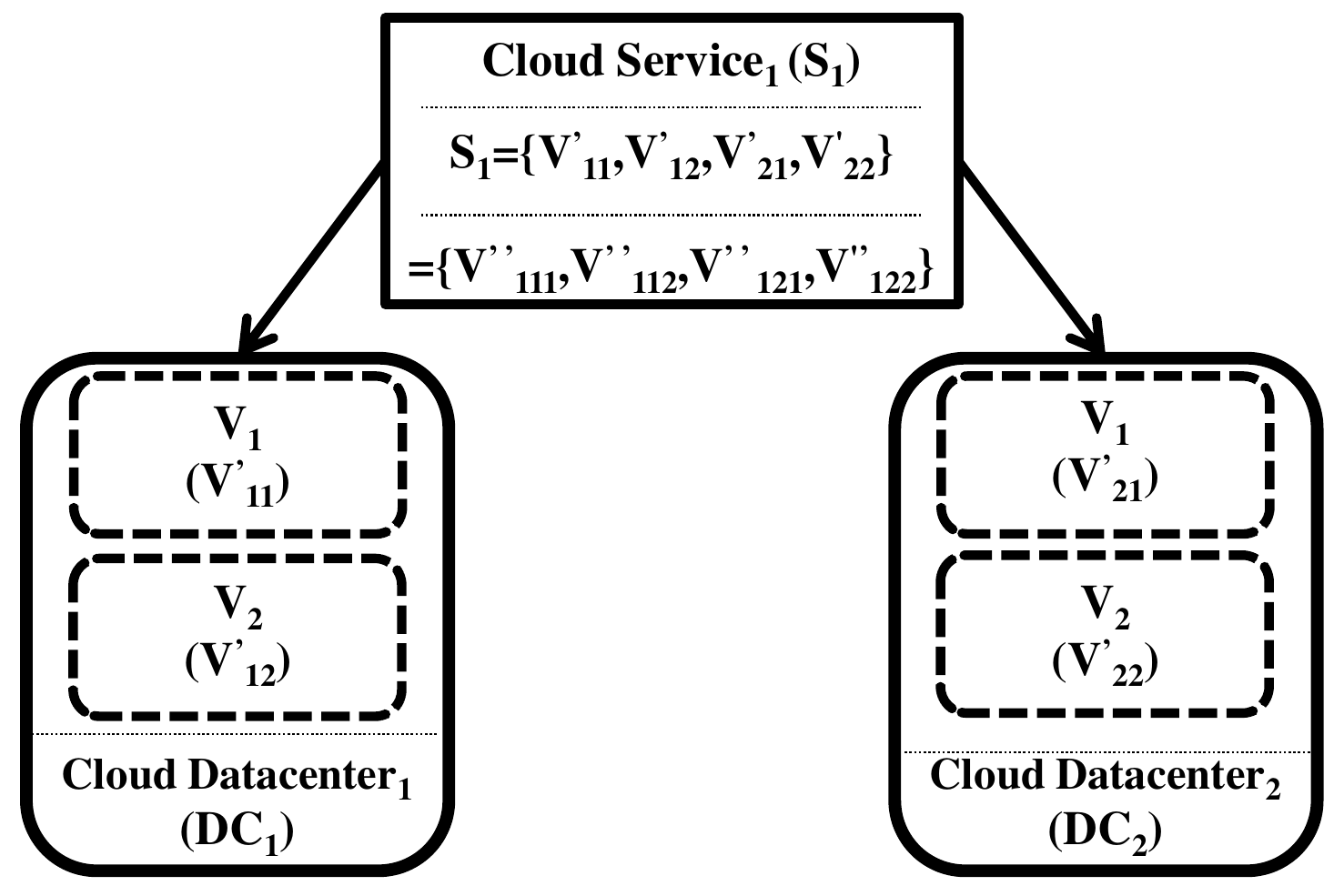}
\caption{Example of a cloud service considered in this work.}
\label{fig:service}
\end{figure}

Complementing the above notation, each cloud datacenter $DC_c$ may be represented as:

\begin{equation}
\label{eq_1}  
DC_{c}=\{V'_{c1},V'_{c2},\dots,V'_{cmDC_c}\}
\end{equation}

\begin{equation}
\label{eq_2}  
\begin{aligned}
V''_{bcj}=&\{Vcpu''_{bcj},Vram''_{bcj},Vnet''_{bcj},R''_{bcj},\\
&SLA''_{bcj},t_{init},t_{end}\}
\end{aligned}
\end{equation}

\noindent where:

\begin{tabbing}
\hspace*{1.65cm}\=\kill
	$V''_{bcj}$:		\>$V_j$ in $DC_{c}$ from service $S_{b}$;\\
	$Vcpu''_{bcj}$:	\>Processing requirements of $V''_{bcj}$ in [ECU];\\
	$Vram''_{bcj}$:	\>Memory requirements of $V''_{bcj}$ in [GB];\\	
	$Vnet''_{bcj}$:	\>Network requirements of $V''_{bcj}$ in [Mbps];\\
	$R''_{bcj}$:		\>Economical revenue for locating $V''_{bcj}$ in [\$];\\
	$SLA''_{bcj}$:	\>SLA of $V''_{bcj}$. $SLA''_{bcj}$ $\in$ $\{1,..,s\}$;\\
	$s$: 						\>Highest priority level of SLAs;\\
	$t_{init}$: 		\>Initial discrete time when $V''_{bcj}$ is executed;\\
	$t_{end}$:			\>Final discrete time when $V''_{bcj}$ is executed.\\
\end{tabbing}
	
Utilization of the resources of each $V''_{bcj}$ is represented by:

\begin{equation}
\label{eq_3}  
U''_{bcj}=\{Ucpu''_{bcj},Uram''_{bcj},Unet''_{bcj}\}
\end{equation}
	
\noindent where:

\begin{tabbing}
\hspace*{1.65cm}\=\kill
	$U''_{bcj}$:		\>Utilization of requirement $V''_{bcj}$;\\
	$Ucpu''_{bcj}$:	\>Utilization of $Vcpu''_{bcj}$ in [ECU]; \\
	$Uram''_{bcj}$:	\>Utilization of $Vram''_{bcj}$ in [GB];\\	
	$Unet''_{bcj}$:	\>Utilization of $Vnet''_{bcj}$ in [Mbps].\\
\end{tabbing}

Note that in practical applications $U''_{bcj}$ is lower than $V''_{bcj}$, giving place to overbooking of resources.

Each of the 16 identified environments considers different parameters that dynamically change as a  function of time $t$, giving place to possible different notations for each environment as presented in Table \ref{table_dynamic_notation}.

\begin{table}[t]
\centering
\caption{Summary of time variables for 16 dynamic environments.}
\label{table_dynamic_notation}
\begin{tabular}{|l|l|l|}
\hline
\bfseries ID & \bfseries Elasticity Type & \bfseries Overbooking Type \\
\hline
(0,0) & Not Considered & Not Considered \\ 
\hline
(0,1) & Not Considered & Server \\
\hline
(0,2) & Not Considered & Network \\
\hline
(0,3) & Not Considered & Server and Network \\ 
\hline
(1,0) & Horizontal & Not Considered \\
\hline
(1,1) & Horizontal & Server \\
\hline
(1,2) & Horizontal & Network \\
\hline
(1,3) & Horizontal & Server and Network \\ 
\hline
(2,0) & Vertical & Not Considered \\
\hline
(2,1) & Vertical & Server \\
\hline
(2,2) & Vertical & Network \\
\hline
(2,3) & Vertical & Server and Network \\ 
\hline
(3,0) & Horizontal and Vertical & Not Considered \\
\hline
(3,1) & Horizontal and Vertical & Server \\
\hline
(3,2) & Horizontal and Vertical & Network \\
\hline
(3,3) & Horizontal and Vertical & Server and Network \\ 
\hline
\end{tabular}
\end{table}

\subsection{Dynamic Environment Examples}
This section will show representative examples with a detailed explanation of 4 of the 16 different dynamic environments.  

Example 1 is presented in Figure \ref{fig:detail_environment_0_1} showing a detailed Environment (0,1). The first level is $CSP_{1}$, it represents a CSP (Cloud Service Provider) requests for allocating VMs. Following, $S_{1}$ shows the requested VMs for the cloud service. $DC_{1}$ and $DC_{2}$ present the resources that the cloud service needs along every discrete time. Because this environment do not considered elasticity, the number of VMs is maintained from time $t=0$ to $t=5$. However, overbooking of server resources is considered and this figure shows how the utilization of CPU and memory varies for every VM in the cloud service. The first VM ($V_{111}$) samples an increase of $Ucpu_{111}$ and $Uram_{111}$ for every discrete time, while $V_{121}$ shows the decrease of $Ucpu_{121}$ from $t=3$ to $t=4$ and an increase of $Uram_{121}$ at the same period time. Other variations of this resources for every VM, executed in this environment, can be observed in this figure.

Example 2 is presented following with the detailed explanation of the environments, Figure \ref{fig:detail_environment_0_2} exposes overbooking of network resources without considering vertical and horizontal elasticity. Analogously to the previews environment, no increase in the number of VMs is presented. A dynamic utilization of network resources is presented in $V_{121}$ with a sharp increase of $Unet_{121}$ from $t=0$ to $t=5$. Other VMs in this environment also present variations of network utilization.

Example 3 Figure \ref{fig:detail_environment_1_0}, shows a detailed example of requirements for Environment (1,0). Knowing that this environment takes into account horizontal elasticity, the table shows the variation of the request of VMs along $t=0$ and $t=5$. From $t=0$ to $t=1$, $CSP_{1}$ executes only four VMs. At the next time unit, the request that the CSP receive changes from two VMs to four VMs. The figure shows that at $t=2$, both datacenters ($DC_{1}$ and $DC_{2}$) needed another resources, for this reason, $S_{2}$ was created with one VM per each datacenter ($V_{3}$ for $DC_{1}$ and $V_{3}$ for $DC_{2}$). This example shows the resources and requirements of each VM that are executed in this environment. The mentioned scenario do not considered overbooking and vertical elasticity, for this reason, the VM requirements and utilization resources stays equal from $t=0$ to $t=5$.

Example 4 The last example, Figure \ref{fig:detail_environment_2_0}, shows a dynamic environment that considers vertical elasticity and no overbooking, Environment (2,0). For $V_{112}$, the requirements $Vcpu_{112}$ and $Vram_{112}$ increase in order to cover the needs of the cloud service. Every VM, detailed in this figure, shows the variation that the requirements present along $t=0$ and $t=5$. 

The 12 remaining environments are composed by the combination of this 4 previously explained environments. Those 12 environments were not detailed with examples due to similarity and content with the previously explained environments.   

\section{Conclusions and Future Directions}

\bibliographystyle{IEEEtranS}
\bibliography{Taxonomy}


\begin{figure*}[ht]
\centering
\renewcommand{\arraystretch}{1.2}
\begin{tabular}{m{1.0cm}:m{2.6cm}:m{2.6cm}:m{2.6cm}:m{2.6cm}:m{2.6cm}:m{0.3cm}}
\cdashline{1-7}
\centering
\bfseries $CSP_{1}$ &
\scriptsize  \bfseries \color{blue} $V''_{111}$, $V''_{112}$, $V''_{121}$, $V''_{122}$ & 
\scriptsize  \bfseries \color{blue} $V''_{111}$, $V''_{112}$, $V''_{121}$, $V''_{122}$ & 
\scriptsize  \bfseries \color{blue} $V''_{111}$, $V''_{112}$, $V''_{121}$, $V''_{122}$ &
\scriptsize  \bfseries \color{blue} $V''_{111}$, $V''_{112}$, $V''_{121}$, $V''_{122}$ &
\scriptsize  \bfseries \color{blue} $V''_{111}$, $V''_{112}$, $V''_{121}$, $V''_{122}$ & \
\\
\cdashline{1-7}
\centering
\bfseries $S_1$ &
\scriptsize  \bfseries \color{blue} $V'_{11}$, $V'_{12}$, $V'_{21}$, $V'_{22}$ & 
\scriptsize  \bfseries \color{blue} $V'_{11}$, $V'_{12}$, $V'_{21}$, $V'_{22}$ & 
\scriptsize  \bfseries \color{blue} $V'_{11}$, $V'_{12}$, $V'_{21}$, $V'_{22}$ & 
\scriptsize  \bfseries \color{blue} $V'_{11}$, $V'_{12}$, $V'_{21}$, $V'_{22}$ & 
\scriptsize  \bfseries \color{blue} $V'_{11}$, $V'_{12}$, $V'_{21}$, $V'_{22}$ & \
\\

\cdashline{1-7}
\centering
\bfseries $DC_1$ &
\scriptsize \bfseries \color{blue} $V_{1}$, $V_{2}$ &  
\scriptsize \bfseries \color{blue} $V_{1}$, $V_{2}$ &  
\scriptsize \bfseries \color{blue} $V_{1}$, $V_{2}$ &  
\scriptsize \bfseries \color{blue} $V_{1}$, $V_{2}$ &  
\scriptsize \bfseries \color{blue} $V_{1}$, $V_{2}$ \
\\ 

\cdashline{1-7}
\centering
\bfseries $DC_2$ &
\scriptsize \bfseries \color{blue} $V_{1}$, $V_{2}$ & 
\scriptsize \bfseries \color{blue} $V_{1}$, $V_{2}$ & 
\scriptsize \bfseries \color{blue} $V_{1}$, $V_{2}$ & 
\scriptsize \bfseries \color{blue} $V_{1}$, $V_{2}$ & 
\scriptsize \bfseries \color{blue} $V_{1}$, $V_{2}$ & 
\scriptsize \bfseries  \
\\

\cdashline{1-7}
\centering
\bfseries $Events$ &
\bfseries \centering \color{ForestGreen} $+S_1$ &
\bfseries $ $ &
\bfseries \centering \color{ForestGreen} $ $ &
\bfseries $ $ &
\bfseries \centering \color{red}  &
\bfseries \color{red} $-S_1$ \
\\

\hline \hline \multicolumn{6}{c}{$V_{111}$}
\\
\hline \hline

\centering
\footnotesize \bfseries $Vcpu_{111}$ &
\bfseries $8$ &
\bfseries $8$ &
\bfseries $8$ &
\bfseries $8$ &
\bfseries  &
\bfseries  \
\\

\cdashline{1-7}
\centering
\footnotesize \bfseries $Vram_{111}$ &
\bfseries $16$ &
\bfseries $16$ &
\bfseries $16$ &
\bfseries $16$ &
\bfseries  &
\bfseries  \
\\

\cdashline{1-7}
\centering
\footnotesize \bfseries $Ucpu_{111}$ &
\bfseries $8$ &
\bfseries $9$ &
\bfseries $10$ &
\bfseries $12$ &
\bfseries  &
\bfseries  \
\\

\cdashline{1-7}
\centering
\footnotesize \bfseries $Uram_{111}$ &
\bfseries $16$ &
\bfseries $18$ &
\bfseries $22$ &
\bfseries $26$ &
\bfseries  &
\bfseries  \
\\

\cdashline{1-7}
\centering
\footnotesize \bfseries $Unet_{111}$ &
\bfseries $150$ &
\bfseries $150$ &
\bfseries $150$ &
\bfseries $150$ &
\bfseries  &
\bfseries  \
\\

\hline \hline \multicolumn{6}{c}{$V_{112}$} \\ \hline \hline

\centering
\footnotesize \bfseries $Vcpu_{112}$ &
\bfseries $5$ &
\bfseries $5$ &
\bfseries $5$ &
\bfseries $5$ &
\bfseries  &
\bfseries  \
\\

\cdashline{1-7}
\centering
\footnotesize \bfseries $Vram_{112}$ &
\bfseries $12$ &
\bfseries $12$ &
\bfseries $12$ &
\bfseries $12$ &
\bfseries  &
\bfseries  \
\\

\cdashline{1-7}
\centering
\footnotesize \bfseries $Ucpu_{112}$ &
\bfseries $5$ &
\bfseries $5$ &
\bfseries $7$ &
\bfseries $7$ &
\bfseries  &
\bfseries  \
\\

\cdashline{1-7}
\centering
\footnotesize \bfseries $Uram_{121}$ &
\bfseries $12$ &
\bfseries $13$ &
\bfseries $16$ &
\bfseries $20$ &
\bfseries  &
\bfseries  \
\\

\cdashline{1-7}
\centering
\footnotesize \bfseries $Unet_{112}$ &
\bfseries $50$ &
\bfseries $50$ &
\bfseries $50$ &
\bfseries $50$ &
\bfseries  &
\bfseries  \
\\


\hline \hline \multicolumn{6}{c}{$V_{121}$} \\ \hline \hline

\centering
\footnotesize \bfseries $Vcpu_{121}$ &
\bfseries $5$ &
\bfseries $5$ &
\bfseries $5$ &
\bfseries $5$ &
\bfseries  &
\bfseries  \
\\

\cdashline{1-7}
\centering
\footnotesize \bfseries $Vram_{121}$ &
\bfseries $12$ &
\bfseries $12$ &
\bfseries $12$ &
\bfseries $12$ &
\bfseries  &
\bfseries  \
\\

\cdashline{1-7}
\centering
\footnotesize \bfseries $Ucpu_{121}$ &
\bfseries $5$ &
\bfseries $4$ &
\bfseries $4$ &
\bfseries $5$ &
\bfseries  &
\bfseries  \
\\

\cdashline{1-7}
\centering
\footnotesize \bfseries $Uram_{121}$ &
\bfseries $12$ &
\bfseries $10$ &
\bfseries $10$ &
\bfseries $9$ &
\bfseries  &
\bfseries  \
\\

\cdashline{1-7}
\centering
\footnotesize \bfseries $Unet_{121}$ &
\bfseries $50$ &
\bfseries $50$ &
\bfseries $50$ &
\bfseries $50$ &
\bfseries  &
\bfseries  \
\\


\hline \hline \multicolumn{6}{c}{$V_{122}$} \\ \hline \hline

\centering
\footnotesize \bfseries $Vcpu_{122}$ &
\bfseries $9$ &
\bfseries $9$ &
\bfseries $9$ &
\bfseries $9$ &
\bfseries  &
\bfseries  \
\\

\cdashline{1-7}
\centering
\footnotesize \bfseries $Vram_{122}$ &
\bfseries $18$ &
\bfseries $18$ &
\bfseries $18$ &
\bfseries $18$ &
\bfseries  &
\bfseries  \
\\

\cdashline{1-7}
\centering
\footnotesize \bfseries $Ucpu_{122}$ &
\bfseries $9$ &
\bfseries $12$ &
\bfseries $14$ &
\bfseries $13$ &
\bfseries  &
\bfseries  \
\\

\cdashline{1-7}
\centering
\footnotesize \bfseries $Uram_{122}$ &
\bfseries $18$ &
\bfseries $20$ &
\bfseries $22$ &
\bfseries $25$ &
\bfseries  &
\bfseries  \
\\

\cdashline{1-7}
\centering
\footnotesize \bfseries $Unet_{122}$ &
\bfseries $170$ &
\bfseries $170$ &
\bfseries $170$ &
\bfseries $170$ &
\bfseries  &
\bfseries  \
\\
\end{tabular}

\begin{tikzpicture}
\draw[very thick,->] (0,0) -- (17.8,0);
\end{tikzpicture}

\begin{tabular}{m{1.0cm}m{2.6cm}m{2.6cm}m{2.6cm}m{2.6cm}m{2.6cm}m{0.3cm}m{0.3cm}}
\centering
\bfseries & 0 & 1 & 2 & 3 & 4 & 5 & (\textit{t})\\ 
\end{tabular}
\caption{Detailed example of dynamic Environment (1,0) for 5 instants of time ($t$)}
\label{fig:detail_environment_0_1}
\end{figure*}


\begin{figure*}[ht]
\centering
\renewcommand{\arraystretch}{1.2}
\begin{tabular}{m{1.0cm}:m{2.6cm}:m{2.6cm}:m{2.6cm}:m{2.6cm}:m{2.6cm}:m{0.3cm}}
\cdashline{1-7}
\centering
\bfseries $CSP_{1}$ &
\scriptsize  \bfseries \color{blue} $V''_{111}$, $V''_{112}$, $V''_{121}$, $V''_{122}$ & 
\scriptsize  \bfseries \color{blue} $V''_{111}$, $V''_{112}$, $V''_{121}$, $V''_{122}$ & 
\scriptsize  \bfseries \color{blue} $V''_{111}$, $V''_{112}$, $V''_{121}$, $V''_{122}$ &
\scriptsize  \bfseries \color{blue} $V''_{111}$, $V''_{112}$, $V''_{121}$, $V''_{122}$ &
\scriptsize  \bfseries \color{blue} $V''_{111}$, $V''_{112}$, $V''_{121}$, $V''_{122}$ & \
\\
\cdashline{1-7}
\centering
\bfseries $S_1$ &
\scriptsize  \bfseries \color{blue} $V'_{11}$, $V'_{12}$, $V'_{21}$, $V'_{22}$ & 
\scriptsize  \bfseries \color{blue} $V'_{11}$, $V'_{12}$, $V'_{21}$, $V'_{22}$ & 
\scriptsize  \bfseries \color{blue} $V'_{11}$, $V'_{12}$, $V'_{21}$, $V'_{22}$ & 
\scriptsize  \bfseries \color{blue} $V'_{11}$, $V'_{12}$, $V'_{21}$, $V'_{22}$ & 
\scriptsize  \bfseries \color{blue} $V'_{11}$, $V'_{12}$, $V'_{21}$, $V'_{22}$ & \
\\

\cdashline{1-7}
\centering
\bfseries $DC_1$ &
\scriptsize \bfseries \color{blue} $V_{1}$, $V_{2}$ &  
\scriptsize \bfseries \color{blue} $V_{1}$, $V_{2}$ &  
\scriptsize \bfseries \color{blue} $V_{1}$, $V_{2}$ &  
\scriptsize \bfseries \color{blue} $V_{1}$, $V_{2}$ &  
\scriptsize \bfseries \color{blue} $V_{1}$, $V_{2}$ \
\\ 

\cdashline{1-7}
\centering
\bfseries $DC_2$ &
\scriptsize \bfseries \color{blue} $V_{1}$, $V_{2}$ & 
\scriptsize \bfseries \color{blue} $V_{1}$, $V_{2}$ & 
\scriptsize \bfseries \color{blue} $V_{1}$, $V_{2}$ & 
\scriptsize \bfseries \color{blue} $V_{1}$, $V_{2}$ & 
\scriptsize \bfseries \color{blue} $V_{1}$, $V_{2}$ & 
\scriptsize \bfseries  \
\\

\cdashline{1-7}
\centering
\bfseries $Events$ &
\bfseries \centering \color{ForestGreen} $+S_1$ &
\bfseries $ $ &
\bfseries \centering \color{ForestGreen} $ $ &
\bfseries $ $ &
\bfseries \centering \color{red}  &
\bfseries \color{red} $-S_1$ \
\\

\hline \hline \multicolumn{6}{c}{$V_{111}$}
\\
\hline \hline

\centering
\footnotesize \bfseries $Vcpu_{111}$ &
\bfseries $8$ &
\bfseries $8$ &
\bfseries $8$ &
\bfseries $8$ &
\bfseries  &
\bfseries  \
\\

\cdashline{1-7}
\centering
\footnotesize \bfseries $Vram_{111}$ &
\bfseries $16$ &
\bfseries $16$ &
\bfseries $16$ &
\bfseries $16$ &
\bfseries  &
\bfseries  \
\\

\cdashline{1-7}
\centering
\footnotesize \bfseries $Ucpu_{111}$ &
\bfseries $8$ &
\bfseries $8$ &
\bfseries $8$ &
\bfseries $8$ &
\bfseries  &
\bfseries  \
\\

\cdashline{1-7}
\centering
\footnotesize \bfseries $Uram_{111}$ &
\bfseries $16$ &
\bfseries $16$ &
\bfseries $16$ &
\bfseries $16$ &
\bfseries  &
\bfseries  \
\\

\cdashline{1-7}
\centering
\footnotesize \bfseries $Unet_{111}$ &
\bfseries $150$ &
\bfseries $170$ &
\bfseries $180$ &
\bfseries $200$ &
\bfseries  &
\bfseries  \
\\

\hline \hline \multicolumn{6}{c}{$V_{112}$} \\ \hline \hline

\centering
\footnotesize \bfseries $Vcpu_{112}$ &
\bfseries $5$ &
\bfseries $5$ &
\bfseries $5$ &
\bfseries $5$ &
\bfseries  &
\bfseries  \
\\

\cdashline{1-7}
\centering
\footnotesize \bfseries $Vram_{112}$ &
\bfseries $12$ &
\bfseries $12$ &
\bfseries $12$ &
\bfseries $12$ &
\bfseries  &
\bfseries  \
\\

\cdashline{1-7}
\centering
\footnotesize \bfseries $Ucpu_{112}$ &
\bfseries $5$ &
\bfseries $5$ &
\bfseries $5$ &
\bfseries $5$ &
\bfseries  &
\bfseries  \
\\

\cdashline{1-7}
\centering
\footnotesize \bfseries $Uram_{121}$ &
\bfseries $12$ &
\bfseries $12$ &
\bfseries $12$ &
\bfseries $12$ &
\bfseries  &
\bfseries  \
\\

\cdashline{1-7}
\centering
\footnotesize \bfseries $Unet_{112}$ &
\bfseries $50$ &
\bfseries $60$ &
\bfseries $40$ &
\bfseries $0$ &
\bfseries  &
\bfseries  \
\\


\hline \hline \multicolumn{6}{c}{$V_{121}$} \\ \hline \hline

\centering
\footnotesize \bfseries $Vcpu_{121}$ &
\bfseries $5$ &
\bfseries $5$ &
\bfseries $5$ &
\bfseries $5$ &
\bfseries  &
\bfseries  \
\\

\cdashline{1-7}
\centering
\footnotesize \bfseries $Vram_{121}$ &
\bfseries $12$ &
\bfseries $12$ &
\bfseries $12$ &
\bfseries $12$ &
\bfseries  &
\bfseries  \
\\

\cdashline{1-7}
\centering
\footnotesize \bfseries $Ucpu_{121}$ &
\bfseries $5$ &
\bfseries $5$ &
\bfseries $5$ &
\bfseries $5$ &
\bfseries  &
\bfseries  \
\\

\cdashline{1-7}
\centering
\footnotesize \bfseries $Uram_{121}$ &
\bfseries $12$ &
\bfseries $12$ &
\bfseries $12$ &
\bfseries $12$ &
\bfseries  &
\bfseries  \
\\

\cdashline{1-7}
\centering
\footnotesize \bfseries $Unet_{121}$ &
\bfseries $50$ &
\bfseries $100$ &
\bfseries $150$ &
\bfseries $170$ &
\bfseries  &
\bfseries  \
\\


\hline \hline \multicolumn{6}{c}{$V_{122}$} \\ \hline \hline

\centering
\footnotesize \bfseries $Vcpu_{122}$ &
\bfseries $9$ &
\bfseries $9$ &
\bfseries $9$ &
\bfseries $9$ &
\bfseries  &
\bfseries  \
\\

\cdashline{1-7}
\centering
\footnotesize \bfseries $Vram_{122}$ &
\bfseries $18$ &
\bfseries $18$ &
\bfseries $18$ &
\bfseries $18$ &
\bfseries  &
\bfseries  \
\\

\cdashline{1-7}
\centering
\footnotesize \bfseries $Ucpu_{122}$ &
\bfseries $9$ &
\bfseries $9$ &
\bfseries $9$ &
\bfseries $9$ &
\bfseries  &
\bfseries  \
\\

\cdashline{1-7}
\centering
\footnotesize \bfseries $Uram_{122}$ &
\bfseries $18$ &
\bfseries $18$ &
\bfseries $18$ &
\bfseries $18$ &
\bfseries  &
\bfseries  \
\\

\cdashline{1-7}
\centering
\footnotesize \bfseries $Unet_{122}$ &
\bfseries $150$ &
\bfseries $200$ &
\bfseries $350$ &
\bfseries $800$ &
\bfseries  &
\bfseries  \
\\

\end{tabular}

\begin{tikzpicture}
\draw[very thick,->] (0,0) -- (17.8,0);
\end{tikzpicture}

\begin{tabular}{m{1.0cm}m{2.6cm}m{2.6cm}m{2.6cm}m{2.6cm}m{2.6cm}m{0.3cm}m{0.3cm}}
\centering
\bfseries & 0 & 1 & 2 & 3 & 4 & 5 & (\textit{t})\\ 
\end{tabular}
\caption{Detailed example of dynamic Environment (0,2) for 5 instants of time ($t$)}
\label{fig:detail_environment_0_2}
\end{figure*}


\begin{figure*}[ht]
\centering
\renewcommand{\arraystretch}{1.2}
\begin{tabular}{m{1.0cm}:m{2.6cm}:m{2.6cm}:m{2.6cm}:m{2.6cm}:m{2.6cm}:m{0.3cm}}
\cdashline{1-7}
\centering
\bfseries $CSP_{1}$ &
\scriptsize  \bfseries \color{blue} $V''_{111}$, $V''_{112}$, $V''_{121}$, $V''_{122}$ & 
\scriptsize  \bfseries \color{blue} $V''_{111}$, $V''_{112}$, $V''_{121}$, $V''_{122}$ & 
\scriptsize  \bfseries \color{blue} $V''_{111}$, $V''_{112}$, $V''_{121}$, $V''_{122}$, \color{Bittersweet} $V''_{213}$,  $V''_{223}$ &
\scriptsize  \bfseries \color{blue} $V''_{111}$, $V''_{112}$, $V''_{121}$, $V''_{122}$, \color{Bittersweet} $V''_{213}$,  $V''_{223}$ &
\scriptsize  \bfseries \color{Bittersweet} $V''_{213}$, $V''_{223}$ & \
\\
\cdashline{1-7}
\centering
\bfseries $S_1$ &
\scriptsize  \bfseries \color{blue} $V'_{11}$, $V'_{12}$, $V'_{21}$, $V'_{22}$ & 
\scriptsize  \bfseries \color{blue} $V'_{11}$, $V'_{12}$, $V'_{21}$, $V'_{22}$ & 
\scriptsize  \bfseries \color{blue} $V'_{11}$, $V'_{12}$, $V'_{21}$, $V'_{22}$ & 
\scriptsize  \bfseries \color{blue} $V'_{11}$, $V'_{12}$, $V'_{21}$, $V'_{22}$ & 
\scriptsize  \bfseries \
\\

\cdashline{1-7}
\centering
\bfseries $S_2$ &
\scriptsize  \bfseries &
\scriptsize  \bfseries &
\scriptsize  \bfseries \color{Bittersweet} $V'_{13}$, $V'_{23}$ & 
\scriptsize  \bfseries \color{Bittersweet} $V'_{13}$, $V'_{23}$ & 
\scriptsize  \bfseries \color{Bittersweet} $V'_{13}$, $V'_{23}$ & 
\\

\cdashline{1-7}
\centering
\bfseries $DC_1$ &
\scriptsize \bfseries \color{blue} $V_{1}$, $V_{2}$ &  
\scriptsize \bfseries \color{blue} $V_{1}$, $V_{2}$ &  
\scriptsize \bfseries \color{blue} $V_{1}$, $V_{2}$, \color{Bittersweet} $V_{3}$ &  
\scriptsize \bfseries \color{blue} $V_{1}$, $V_{2}$, \color{Bittersweet} $V_{3}$ &  
\scriptsize \bfseries \color{Bittersweet} $V_{3}$ \
\\ 

\cdashline{1-7}
\centering
\bfseries $DC_2$ &
\scriptsize \bfseries \color{blue} $V_{1}$, $V_{2}$ & 
\scriptsize \bfseries \color{blue} $V_{1}$, $V_{2}$ & 
\scriptsize \bfseries \color{blue} $V_{1}$, $V_{2}$, \color{Bittersweet} $V_{3}$ & 
\scriptsize \bfseries \color{blue} $V_{1}$, $V_{2}$, \color{Bittersweet} $V_{3}$ & 
\scriptsize \bfseries \color{Bittersweet} $V_{3}$ &
\scriptsize \bfseries \
\\

\cdashline{1-7}
\centering
\bfseries $Events$ &
\bfseries \centering \color{ForestGreen} $+S_1$ &
\bfseries $ $ &
\bfseries \centering \color{ForestGreen} $+S_2$ &
\bfseries $ $ &
\bfseries \centering \color{red} $-S_1$ &
\bfseries \color{red} $-S_2$ \
\\

\hline \hline \multicolumn{6}{c}{$V_{111}$}
\\
\hline \hline

\centering
\footnotesize \bfseries $Vcpu_{111}$ &
\bfseries $8$ &
\bfseries $8$ &
\bfseries $8$ &
\bfseries $8$ &
\bfseries  &
\bfseries  \
\\

\cdashline{1-7}
\centering
\footnotesize \bfseries $Vram_{111}$ &
\bfseries $16$ &
\bfseries $16$ &
\bfseries $16$ &
\bfseries $16$ &
\bfseries  &
\bfseries  \
\\

\cdashline{1-7}
\centering
\footnotesize \bfseries $Ucpu_{111}$ &
\bfseries $8$ &
\bfseries $8$ &
\bfseries $8$ &
\bfseries $8$ &
\bfseries  &
\bfseries  \
\\

\cdashline{1-7}
\centering
\footnotesize \bfseries $Uram_{111}$ &
\bfseries $16$ &
\bfseries $16$ &
\bfseries $16$ &
\bfseries $16$ &
\bfseries  &
\bfseries  \
\\

\cdashline{1-7}
\centering
\footnotesize \bfseries $Unet_{111}$ &
\bfseries $150$ &
\bfseries $150$ &
\bfseries $150$ &
\bfseries $150$ &
\bfseries  &
\bfseries  \
\\

\hline \hline \multicolumn{6}{c}{$V_{112}$} \\ \hline \hline

\centering
\footnotesize \bfseries $Vcpu_{112}$ &
\bfseries $5$ &
\bfseries $5$ &
\bfseries $5$ &
\bfseries $5$ &
\bfseries  &
\bfseries  \
\\

\cdashline{1-7}
\centering
\footnotesize \bfseries $Vram_{112}$ &
\bfseries $12$ &
\bfseries $12$ &
\bfseries $12$ &
\bfseries $12$ &
\bfseries  &
\bfseries  \
\\

\cdashline{1-7}
\centering
\footnotesize \bfseries $Ucpu_{112}$ &
\bfseries $5$ &
\bfseries $5$ &
\bfseries $5$ &
\bfseries $5$ &
\bfseries  &
\bfseries  \
\\

\cdashline{1-7}
\centering
\footnotesize \bfseries $Uram_{121}$ &
\bfseries $12$ &
\bfseries $12$ &
\bfseries $12$ &
\bfseries $12$ &
\bfseries  &
\bfseries  \
\\

\cdashline{1-7}
\centering
\footnotesize \bfseries $Unet_{112}$ &
\bfseries $50$ &
\bfseries $50$ &
\bfseries $50$ &
\bfseries $50$ &
\bfseries  &
\bfseries  \
\\


\hline \hline \multicolumn{6}{c}{$V_{121}$} \\ \hline \hline

\centering
\footnotesize \bfseries $Vcpu_{121}$ &
\bfseries $5$ &
\bfseries $5$ &
\bfseries $5$ &
\bfseries $5$ &
\bfseries  &
\bfseries  \
\\

\cdashline{1-7}
\centering
\footnotesize \bfseries $Vram_{121}$ &
\bfseries $12$ &
\bfseries $12$ &
\bfseries $12$ &
\bfseries $12$ &
\bfseries  &
\bfseries  \
\\

\cdashline{1-7}
\centering
\footnotesize \bfseries $Ucpu_{121}$ &
\bfseries $5$ &
\bfseries $5$ &
\bfseries $5$ &
\bfseries $5$ &
\bfseries  &
\bfseries  \
\\

\cdashline{1-7}
\centering
\footnotesize \bfseries $Uram_{121}$ &
\bfseries $12$ &
\bfseries $12$ &
\bfseries $12$ &
\bfseries $12$ &
\bfseries  &
\bfseries  \
\\

\cdashline{1-7}
\centering
\footnotesize \bfseries $Unet_{121}$ &
\bfseries $50$ &
\bfseries $50$ &
\bfseries $50$ &
\bfseries $50$ &
\bfseries  &
\bfseries  \
\\


\hline \hline \multicolumn{6}{c}{$V_{122}$} \\ \hline \hline

\centering
\footnotesize \bfseries $Vcpu_{122}$ &
\bfseries $9$ &
\bfseries $9$ &
\bfseries $9$ &
\bfseries $9$ &
\bfseries  &
\bfseries  \
\\

\cdashline{1-7}
\centering
\footnotesize \bfseries $Vram_{122}$ &
\bfseries $18$ &
\bfseries $18$ &
\bfseries $18$ &
\bfseries $18$ &
\bfseries  &
\bfseries  \
\\

\cdashline{1-7}
\centering
\footnotesize \bfseries $Ucpu_{122}$ &
\bfseries $9$ &
\bfseries $9$ &
\bfseries $9$ &
\bfseries $9$ &
\bfseries  &
\bfseries  \
\\

\cdashline{1-7}
\centering
\footnotesize \bfseries $Uram_{122}$ &
\bfseries $18$ &
\bfseries $18$ &
\bfseries $18$ &
\bfseries $18$ &
\bfseries  &
\bfseries  \
\\

\cdashline{1-7}
\centering
\footnotesize \bfseries $Unet_{122}$ &
\bfseries $170$ &
\bfseries $170$ &
\bfseries $170$ &
\bfseries $170$ &
\bfseries  &
\bfseries  \
\\


\hline \hline \multicolumn{6}{c}{$V_{213}$} \\ \hline \hline

\centering
\footnotesize \bfseries $Vcpu_{213}$ &
\bfseries  &
\bfseries  &
\bfseries $2$ &
\bfseries $2$ &
\bfseries $2$ &
\bfseries  \
\\

\cdashline{1-7}
\centering
\footnotesize \bfseries $Vram_{213}$ &
\bfseries  &
\bfseries  &
\bfseries $10$ &
\bfseries $10$ &
\bfseries $10$ &
\bfseries  \
\\

\cdashline{1-7}
\centering
\footnotesize \bfseries $Ucpu_{213}$ &
\bfseries  &
\bfseries  &
\bfseries $2$ &
\bfseries $2$ &
\bfseries $2$ &
\bfseries  \
\\

\cdashline{1-7}
\centering
\footnotesize \bfseries $Uram_{213}$ &
\bfseries  &
\bfseries  &
\bfseries $10$ &
\bfseries $10$ &
\bfseries $10$ &
\bfseries  \
\\

\cdashline{1-7}
\centering
\footnotesize \bfseries $Unet_{213}$ &
\bfseries  &
\bfseries  &
\bfseries $60$ &
\bfseries $60$ &
\bfseries $60$ &
\bfseries  \
\\

\hline \hline \multicolumn{6}{c}{$V_{223}$} \\ \hline \hline

\centering
\footnotesize \bfseries $Vcpu_{223}$ &
\bfseries  &
\bfseries  &
\bfseries $8$ &
\bfseries $8$ &
\bfseries $8$ &
\bfseries  \
\\

\cdashline{1-7}
\centering
\footnotesize \bfseries $Vram_{223}$ &
\bfseries  &
\bfseries  &
\bfseries $20$ &
\bfseries $20$ &
\bfseries $20$ &
\bfseries  \
\\

\cdashline{1-7}
\centering
\footnotesize \bfseries $Ucpu_{223}$ &
\bfseries  &
\bfseries  &
\bfseries $8$ &
\bfseries $8$ &
\bfseries $8$ &
\bfseries  \
\\

\cdashline{1-7}
\centering
\footnotesize \bfseries $Uram_{223}$ &
\bfseries  &
\bfseries  &
\bfseries $20$ &
\bfseries $20$ &
\bfseries $20$ &
\bfseries  \
\\

\cdashline{1-7}
\centering
\footnotesize \bfseries $Unet_{223}$ &
\bfseries  &
\bfseries  &
\bfseries $120$ &
\bfseries $120$ &
\bfseries $120$ &
\bfseries  \
\\
\end{tabular}

\begin{tikzpicture}
\draw[very thick,->] (0,0) -- (17.8,0);
\end{tikzpicture}

\begin{tabular}{m{1.0cm}m{2.6cm}m{2.6cm}m{2.6cm}m{2.6cm}m{2.6cm}m{0.3cm}m{0.3cm}}
\centering
\bfseries & 0 & 1 & 2 & 3 & 4 & 5 & (\textit{t})\\ 
\end{tabular}
\caption{Detailed example of dynamic Environment (1,0) for 5 instants of time ($t$)}
\label{fig:detail_environment_1_0}
\end{figure*}


\begin{figure*}[ht]
\centering
\renewcommand{\arraystretch}{1.2}
\begin{tabular}{m{1.0cm}:m{2.6cm}:m{2.6cm}:m{2.6cm}:m{2.6cm}:m{2.6cm}:m{0.3cm}}
\cdashline{1-7}
\centering
\bfseries $CSP_{1}$ &
\scriptsize  \bfseries \color{blue} $V''_{111}$, $V''_{112}$, $V''_{121}$, $V''_{122}$ & 
\scriptsize  \bfseries \color{blue} $V''_{111}$, $V''_{112}$, $V''_{121}$, $V''_{122}$ & 
\scriptsize  \bfseries \color{blue} $V''_{111}$, $V''_{112}$, $V''_{121}$, $V''_{122}$ &
\scriptsize  \bfseries \color{blue} $V''_{111}$, $V''_{112}$, $V''_{121}$, $V''_{122}$ &
\scriptsize  \bfseries \color{blue} $V''_{111}$, $V''_{112}$, $V''_{121}$, $V''_{122}$ & \
\\
\cdashline{1-7}
\centering
\bfseries $S_1$ &
\scriptsize  \bfseries \color{blue} $V'_{11}$, $V'_{12}$, $V'_{21}$, $V'_{22}$ & 
\scriptsize  \bfseries \color{blue} $V'_{11}$, $V'_{12}$, $V'_{21}$, $V'_{22}$ & 
\scriptsize  \bfseries \color{blue} $V'_{11}$, $V'_{12}$, $V'_{21}$, $V'_{22}$ & 
\scriptsize  \bfseries \color{blue} $V'_{11}$, $V'_{12}$, $V'_{21}$, $V'_{22}$ & 
\scriptsize  \bfseries \color{blue} $V'_{11}$, $V'_{12}$, $V'_{21}$, $V'_{22}$ & \
\\

\cdashline{1-7}
\centering
\bfseries $DC_1$ &
\scriptsize \bfseries \color{blue} $V_{1}$, $V_{2}$ &  
\scriptsize \bfseries \color{blue} $V_{1}$, $V_{2}$ &  
\scriptsize \bfseries \color{blue} $V_{1}$, $V_{2}$ &  
\scriptsize \bfseries \color{blue} $V_{1}$, $V_{2}$ &  
\scriptsize \bfseries \color{blue} $V_{1}$, $V_{2}$ \
\\ 

\cdashline{1-7}
\centering
\bfseries $DC_2$ &
\scriptsize \bfseries \color{blue} $V_{1}$, $V_{2}$ & 
\scriptsize \bfseries \color{blue} $V_{1}$, $V_{2}$ & 
\scriptsize \bfseries \color{blue} $V_{1}$, $V_{2}$ & 
\scriptsize \bfseries \color{blue} $V_{1}$, $V_{2}$ & 
\scriptsize \bfseries \color{blue} $V_{1}$, $V_{2}$ & 
\scriptsize \bfseries  \
\\

\cdashline{1-7}
\centering
\bfseries $Events$ &
\bfseries \centering \color{ForestGreen} $+S_1$ &
\bfseries $ $ &
\bfseries \centering \color{ForestGreen} $ $ &
\bfseries $ $ &
\bfseries \centering \color{red}  &
\bfseries \color{red} $-S_1$ \
\\

\hline \hline \multicolumn{6}{c}{$V_{111}$}
\\
\hline \hline

\centering
\footnotesize \bfseries $Vcpu_{111}$ &
\bfseries $8$ &
\bfseries $9$ &
\bfseries $9$ &
\bfseries $11$ &
\bfseries  &
\bfseries  \
\\

\cdashline{1-7}
\centering
\footnotesize \bfseries $Vram_{111}$ &
\bfseries $16$ &
\bfseries $22$ &
\bfseries $23$ &
\bfseries $25$ &
\bfseries  &
\bfseries  \
\\

\cdashline{1-7}
\centering
\footnotesize \bfseries $Ucpu_{111}$ &
\bfseries $8$ &
\bfseries $8$ &
\bfseries $8$ &
\bfseries $8$ &
\bfseries  &
\bfseries  \
\\

\cdashline{1-7}
\centering
\footnotesize \bfseries $Uram_{111}$ &
\bfseries $16$ &
\bfseries $16$ &
\bfseries $16$ &
\bfseries $16$ &
\bfseries  &
\bfseries  \
\\

\cdashline{1-7}
\centering
\footnotesize \bfseries $Unet_{111}$ &
\bfseries $150$ &
\bfseries $150$ &
\bfseries $150$ &
\bfseries $150$ &
\bfseries  &
\bfseries  \
\\

\hline \hline \multicolumn{6}{c}{$V_{112}$} \\ \hline \hline

\centering
\footnotesize \bfseries $Vcpu_{112}$ &
\bfseries $5$ &
\bfseries $6$ &
\bfseries $6$ &
\bfseries $5$ &
\bfseries  &
\bfseries  \
\\

\cdashline{1-7}
\centering
\footnotesize \bfseries $Vram_{112}$ &
\bfseries $12$ &
\bfseries $15$ &
\bfseries $15$ &
\bfseries $16$ &
\bfseries  &
\bfseries  \
\\

\cdashline{1-7}
\centering
\footnotesize \bfseries $Ucpu_{112}$ &
\bfseries $5$ &
\bfseries $5$ &
\bfseries $5$ &
\bfseries $5$ &
\bfseries  &
\bfseries  \
\\

\cdashline{1-7}
\centering
\footnotesize \bfseries $Uram_{121}$ &
\bfseries $12$ &
\bfseries $12$ &
\bfseries $12$ &
\bfseries $12$ &
\bfseries  &
\bfseries  \
\\

\cdashline{1-7}
\centering
\footnotesize \bfseries $Unet_{112}$ &
\bfseries $50$ &
\bfseries $50$ &
\bfseries $50$ &
\bfseries $50$ &
\bfseries  &
\bfseries  \
\\


\hline \hline \multicolumn{6}{c}{$V_{121}$} \\ \hline \hline

\centering
\footnotesize \bfseries $Vcpu_{121}$ &
\bfseries $5$ &
\bfseries $6$ &
\bfseries $6$ &
\bfseries $7$ &
\bfseries  &
\bfseries  \
\\

\cdashline{1-7}
\centering
\footnotesize \bfseries $Vram_{121}$ &
\bfseries $12$ &
\bfseries $18$ &
\bfseries $18$ &
\bfseries $20$ &
\bfseries  &
\bfseries  \
\\

\cdashline{1-7}
\centering
\footnotesize \bfseries $Ucpu_{121}$ &
\bfseries $5$ &
\bfseries $5$ &
\bfseries $5$ &
\bfseries $5$ &
\bfseries  &
\bfseries  \
\\

\cdashline{1-7}
\centering
\footnotesize \bfseries $Uram_{121}$ &
\bfseries $12$ &
\bfseries $12$ &
\bfseries $12$ &
\bfseries $12$ &
\bfseries  &
\bfseries  \
\\

\cdashline{1-7}
\centering
\footnotesize \bfseries $Unet_{121}$ &
\bfseries $50$ &
\bfseries $50$ &
\bfseries $50$ &
\bfseries $50$ &
\bfseries  &
\bfseries  \
\\


\hline \hline \multicolumn{6}{c}{$V_{122}$} \\ \hline \hline

\centering
\footnotesize \bfseries $Vcpu_{122}$ &
\bfseries $9$ &
\bfseries $6$ &
\bfseries $6$ &
\bfseries $5$ &
\bfseries  &
\bfseries  \
\\

\cdashline{1-7}
\centering
\footnotesize \bfseries $Vram_{122}$ &
\bfseries $18$ &
\bfseries $10$ &
\bfseries $10$ &
\bfseries $12$ &
\bfseries  &
\bfseries  \
\\

\cdashline{1-7}
\centering
\footnotesize \bfseries $Ucpu_{122}$ &
\bfseries $9$ &
\bfseries $9$ &
\bfseries $9$ &
\bfseries $9$ &
\bfseries  &
\bfseries  \
\\

\cdashline{1-7}
\centering
\footnotesize \bfseries $Uram_{122}$ &
\bfseries $18$ &
\bfseries $18$ &
\bfseries $18$ &
\bfseries $18$ &
\bfseries  &
\bfseries  \
\\

\cdashline{1-7}
\centering
\footnotesize \bfseries $Unet_{122}$ &
\bfseries $170$ &
\bfseries $170$ &
\bfseries $170$ &
\bfseries $170$ &
\bfseries  &
\bfseries  \
\\

\end{tabular}

\begin{tikzpicture}
\draw[very thick,->] (0,0) -- (17.8,0);
\end{tikzpicture}

\begin{tabular}{m{1.0cm}m{2.6cm}m{2.6cm}m{2.6cm}m{2.6cm}m{2.6cm}m{0.3cm}m{0.3cm}}
\centering
\bfseries & 0 & 1 & 2 & 3 & 4 & 5 & (\textit{t})\\ 
\end{tabular}
\caption{Detailed example of dynamic Environment (2,0) for 5 instants of time ($t$)}
\label{fig:detail_environment_2_0}
\end{figure*}

\end{document}